\documentclass[english,notitlepage,twocolumn]{revtex4-2}
\usepackage[utf8]{inputenc}
\usepackage{color}
\usepackage{amsbsy}
\usepackage{amstext}
\usepackage{amsmath}
\usepackage{amssymb}
\usepackage{graphicx}
\usepackage{esint}
\usepackage{setspace} 
\usepackage{wrapfig}
\usepackage[english]{babel}

\begin{document}
\title{Enhanced Enantioselective Optical Trapping enabled by \\ Longitudinal Mie Resonances in Silicon Nanodisks}

\author{Guillermo Serrera}
\email{guillermo.serrera@unican.es}
\author{Pablo Albella}
\email{pablo.albella@unican.es}
\affiliation{%
Group of Optics, Department of Applied Physics, University of Cantabria, 39005, Spain
}%

\begin{abstract}
Optical enantioseparation of nanoscale matter is fundamentally limited by the intrinsic weakness of chiroptical forces compared to the dominant achiral gradient forces and thermal fluctuations. Conventional plasmonic approaches typically enhance chirality at the cost of amplifying achiral attraction and heating. Here, we overcome this trade-off by exploiting longitudinal Mie resonances in silicon nanodisks. By employing an Azimuthally–Radially Polarized Beam (ARPB) illumination, we excite longitudinal Mie resonances, with strong optical chirality gradients and comparatively uniform electric field intensities. Specifically, magnetic quadrupole (MQ) resonances effectively decouple enantioselective forces from the achiral background, providing uniquely favorable conditions for enantioselective optical trapping. Combining numerical simulations with Kramers’ escape-rate theory, we demonstrate a robust and highly selective trapping system that is also experimentally accessible. We predict trapping selectivity ratios above 100 for particles with Pasteur parameters $|\kappa| \geq 0.03$ and maintain selectivities above 2 even for weakly chiral analytes ($|\kappa| \approx 0.006$). These results establish longitudinal Mie resonances in high-index dielectric nanostructures as a promising, non-invasive platform for all-optical chiral analysis and enantiomer separation. 
\end{abstract}

\maketitle

\section{Introduction}

Chirality, the impossibility of superimposing an object onto its mirror image \cite{Pasteur1848, ThomsonBaronKelvin1894, ThomsonBaronKelvin1904}, plays an increasingly important role across different disciplines in the nanosciences, including biochemistry, nanotechnology and nanophotonics \cite{Yoo2019,Zhang2005,Castillo2023}. Given the presence of homochirality in natural systems, their functionality is often dictated by chirality. In particular, although chirality is widely associated with biomolecular recognition and pharmaceutical activity \cite{Dobson2003,Nguyen2006}, recent advances in nanoscience have enabled the synthesis of a broad range of chiral nanoparticles with applications in polarization control, sensing, and asymmetric catalysis \cite{GonzalezRubio2020,Zheng2025,Zafar2020,Adler2015}. Furthermore, chirality also plays a significant role in the functionalization and assembly of different functional nanostructures, including those involving DNA techniques \cite{Winogradoff2021,Tortora2020,Siegel2025,SanzPaz2026}. As a result, the development of optical strategies capable of selectively probing and manipulating chiral nanoscale objects has emerged as a major challenge in modern photonics \cite{Nordn1997}.

In contrast to chemical enantioseparation methods, often limited by their invasive nature and poor selectivity \cite{Sui2023,Lmmerhofer2010}, optical manipulation techniques offer a non-invasive, label-free alternative. Although longstanding methods like Circular Dichroism (CD) and Circularly Polarized Luminescence (CPL) are well established, they suffer from intrinsically weak signals. To overcome this, nanophotonic platforms have been extensively engineered to enhance chiral light-matter interactions \cite{Serrera2024, Mohammadi2021, Mohammadi2023, Raziman2019}.

Beyond mere detection, chiral optical forces ---small, chirality-dependent forces--- have emerged as a mechanism for the passive, mechanical separation of enantiomers \cite{Canaguier-Durand2013,Genet2022}. These forces typically scale with the Pasteur parameter $\kappa$ of the analyte, which quantifies the material’s chiral response to electromagnetic fields; and typically manifest in the form of chirality-dependent gradient or scattering forces. Successful experimental implementations of such forces for enantiomer discrimination have been recently demonstrated, highlighting the growing maturity of this approach \cite{Yamanishi2022,Tkachenko2026}. However, a fundamental bottleneck remains: chiral optical forces are orders of magnitude weaker than their achiral counterparts (e.g., standard gradient and radiation pressure forces), particularly for natural chiral media, where $\kappa$ is very small \cite{Mohammadi2023,Lindell1994}. This limits their effectiveness in practical applications, where long measurement times and precise analyte placement are typically required to observe any chirality-dependent response \cite{Yamane2023}. To mitigate the weakness of chiroptical effects, strategies involving evanescent wave excitation and nanostructured materials or waveguides have been leveraged \cite{Diez2024,Martnez-Romeu2024,Golat2024}.

Specifically, efforts to enhance enantioselective optical trapping via gradient forces have relied on strong local field variations—commonly achieved using evanescent waves or resonant nanostructures \cite{Hayat2015,Li2020,Wang2014}. Surface Plasmon Resonances (SPRs), for example, generate intense electric-field gradients that enhance the Optical Chirality Density (OCD) \cite{Tang2010,Hentschel2017,Lin2021,Zhao2016}. However, this approach presents an inherent trade-off: the strong field gradients required to boost the OCD simultaneously amplify the achiral gradient forces. Consequently, the selectivity of the optical trap degrades, as the achiral background force continues to dominate unless the analyte possesses an unrealistically high $\kappa$ value. Furthermore, the high optical intensities required for optical trapping may induce thermoplasmonic effects \cite{Govorov2007,Gonzlez-Colsa2022}, which may compromise stability or damage delicate biological analytes. These limitations highlight the need for optical platforms capable of enhancing chiral forces over achiral ones, leveraging OCD gradients over high electric field enhancements \cite{Schferling2018}. 

In this work, we propose high-refractive-index silicon nanodisks to overcome these limitations. By illuminating these nanostructures with Azimuthally-Radially Polarized Beams (ARPB), we excite longitudinal Mie resonances, generating strong OCD gradients on the upper faces of disks. Crucially, these gradients are driven by magnetic rather than electric field enhancements. This mechanism minimizes photothermal heating while maximizing the chiral-to-achiral force ratio, enabling reliable enantioselection even for analytes with modest $\kappa$ values. Following optimization of the geometrical and illumination parameters, we employ Kramers’ escape rate theory to quantify the stability of the enantiomer-selective optical trap. Notably, our results demonstrate theoretical trap selectivities exceeding 100 for $\kappa \geq 0.03$ and maintain selectivities above 2 even for weakly chiral particles with $\kappa < 0.01$.

\section{Longitudinal Mie resonances and optical chirality in \\ silicon nanodisks}

To realize such enantioselective chiral forces, we exploit Azimuthally-Radially Polarized Beams (ARPBs) \cite{Herrero-Parareda2024,Herrero-Parareda2025}. These structured fields are formed by the coherent superposition of an azimuthally polarized beam (APB) and a radially polarized beam (RPB), with a constant phase difference $\Psi$. The total electric field distribution can be expressed as:

\begin{multline}
    \label{eq:1}
    \mathbf{E^{ARPB}} = \mathbf{E^{APB}} + \mathbf{E^{RPB}} = \frac{V_R}{w^2} f(A_\rho + i B_\rho) \boldsymbol{\hat{\rho}} + \\ + \frac{V_A}{w^2}f e^{i\Psi} \boldsymbol{\hat{\varphi}} + \frac{2iV_R}{kw^2} f(A_z + iB_z) \mathbf{\hat{z}}
\end{multline}
where $V_R$ and $V_A$ denote amplitude coefficients for the radial and azimuthal components, respectively; $f$ represents a Gaussian beam envelope function, characterized by radius $w$; and the $A$ and $B$ parameters being shorthand for the full expression, provided in the Supporting Information (appendix A). As revealed by equation \ref{eq:1}, both the electric and magnetic fields have strong longitudinal field components along the beam axis (with RPB contributing a strong longitudinal electric field $E_z$ and the APB a magnetic field $H_z$). Due to the dual nature of the APB and RPB beams, when $V_R = V_A$ and $\Psi = \pm \pi/2$, the eelectric and magnetic fields align and maintain a global $\pm \pi/2$ phase difference. This means that the optical field becomes optimally chiral, maximizing the OCD $C= -\frac{\omega}{2c^2}\mathfrak{Im}(\mathbf{E^*} \cdot \mathbf{H})$ for a given energy density, a condition also met in circularly polarized light \cite{Hanifeh2020} and engineered in other nanophotonic devices for chiral sensing \cite{Mohammadi2021,Tian2015,Yao2018}. Furthermore, the presence of strong longitudinal fields can result in the excitation of longitudinal Mie resonances in nanostructures.

Small chiral particles can be modeled as a pair of coupled electric ($\mathbf{p}$) and magnetic ($\mathbf{m}$) point dipoles. For such particles, the time-averaged optical force, under electric and magnetic fields $\mathbf{E}$ and $\mathbf{H}$ at frequency $\omega$, is given by \cite{Chantada2010}

\begin{multline}
\label{eq:2}
    \langle \mathbf{F} \rangle = \frac{1}{2} \mathfrak{Re}\left[ (\nabla \mathbf{E^*})\cdot \mathbf{p} + (\nabla \mathbf{H^*}) \cdot \mathbf{m} - \right. \\ \left. - \frac{ck^4}{6\pi} (\mathbf{p} \times \mathbf{m^*})\right]
\end{multline}
where the dipole moments are defined via the polarizabilities: $\mathbf{p} = \alpha_e \mathbf{E} + i \alpha_{em} \mathbf{H}$ and $\mathbf{m} = \alpha_m \mathbf{H}- i \alpha_{em} \mathbf{E}$. Here, $\alpha_e$ and $\alpha_m$ represent the electric and magnetic polarizabilities, while the electromagnetic polarizability $\alpha_{em}$ mediates the interaction between the two dipoles. For small, low-loss chiral spheres, operating off-resonance, one may take $\alpha_e \approx \mathfrak{Re}(\alpha_e)$, $\alpha_m \approx 0$, and $\alpha_{em} \approx \mathfrak{Re}(\alpha_{em})$; which simplifies the force expression considerably \cite{Wang2014}:

\begin{multline}
\label{eq:3}
    \langle \mathbf{F} \rangle \approx \frac{1}{4} \mathfrak{Re}(\alpha_e) \nabla |\mathbf{E}|^2 - \frac{1}{2} \mathfrak{Re}(\alpha_{em}) \nabla \left[\mathfrak{Im}(\mathbf{E^*} \cdot \mathbf{H})\right] + \\ + \sigma \frac{\langle \mathbf{S} \rangle}{c} + \omega \gamma_e \langle \mathbf{L_e} \rangle 
\end{multline}
where $\langle \mathbf{S} \rangle$ is the time-averaged Poynting vector, $\langle \mathbf{L_e} \rangle = \frac{\varepsilon_0}{4\omega i}(\mathbf{E} \times \mathbf{E^*})$  is the time-averaged optical spin density, $\sigma = - \frac{c^2 k_0^4}{6\pi}(\alpha_{em} \alpha_{em}^*)$, and $\gamma_e \approx \frac{ck^4}{3\pi\varepsilon_0}\mathfrak{Re}(\alpha_e \alpha_{em}^*)$. This expression contains both reactive (gradient-dependent) and dissipative (scattering and spin) terms. As a result of considering real polarizabilities, beyond the molecular-resonant spectral regions, dissipative terms are typically negligible in comparison with the reactive terms.

Thus, in this regime, and excluding dissipative terms, the time-averaged optical force is dominated by the gradients of both the electric field intensity and $\mathfrak{Im}(\mathbf{E^*} \cdot \mathbf{H})$ (which defines the shape of the OCD). Successful enantioselective trapping, therefore, relies on maximizing the latter while preventing the achiral contribution from dominating the total force.

\begin{figure*}
    \includegraphics[width=\textwidth]{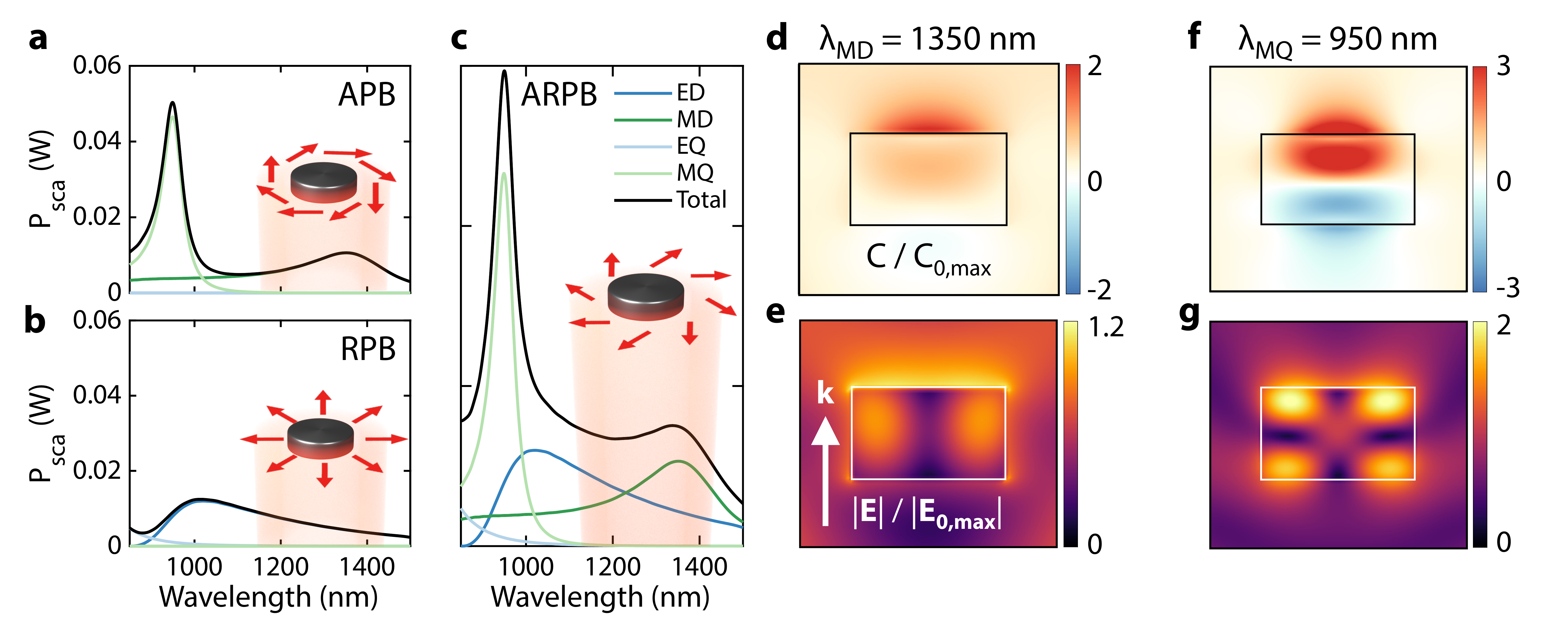}
    \caption{Scattering of azimuthally and radially polarized beams by a silicon nanodisk with radius $R$ = 180 nm and height $h$ = 240 nm. a-c) Scattered power upon illumination with a 1W power a) Azimuthally Polarized Beam (APB), b) Radially Polarized Beam (RPB), and c) Azimuthally-Radially Polarized Beam (ARPB). d) Normalized OCD and e) electric field magnitude in the nanodisk vicinity at the Magnetic Dipole (MD) resonance and f-g) Magnetic Quadrupole (MQ) resonance. Normalizations are performed considering the points of maximum intensity of the incident beams.}
    \label{fig:1}
\end{figure*}

To investigate the capacity of longitudinal Mie resonances in silicon nanodisks to provide such gradients, we perform simulations of single silicon nanodisks (radius $R$ = 180 nm, height $h$ = 240 nm) in water. The structures were illuminated separately by APB and RPB sources (waist radius $w_0$ equal to the wavelength $\lambda$). A high beam power of 1 W was used to provide an easier interpretation of simulation results. As shown in Figure \ref{fig:1}a and b, respectively, these beams excite longitudinal resonances of either magnetic or electric character, depending on the beam polarization. Consequently, the ARPB combination (Figure \ref{fig:1}c), supports the simultaneous excitation of both electric and magnetic resonances, a crucial feature for enhanced chiral light-matter interactions. A multipole decomposition reveals a prominent magnetic quadrupole (MQ) resonance near 950 nm, and a broader and weaker resonance from the magnetic dipole (MD) at 1350 nm. A cartesian component analysis of these resonances (see Supporting Information, appendix B) confirms that indeed the resonances are fundamentally longitudinal.

Next, we examine the near-fields around the nanodisks at the magnetic resonances (Figure \ref{fig:1}d-g), where strong optical chirality is expected. In the vicinity of the MD resonance (1350 nm), the OCD is enhanced by a factor up to 1.6 at the center of the disk’s upper face, whereas the electric field magnitude remains relatively spatially uniform. This contrast effectively supports the potential for enhanced chiral force contributions in this platform.

This idea is also supported by fields close to the magnetic quadrupole resonance (950 nm), where the effect is further amplified. Both the magnetic and electric fields become stronger, with the electric field forming an annular (ring-like) profile on the disk surface. Despite this local enhancement, the overall electric field profile preserves sufficient uniformity to prevent achiral forces from overwhelming the trap, thereby maintaining a favorable landscape for enantioselective optical trapping.

\begin{figure*}
    \centering
    \includegraphics[width=0.8\textwidth]{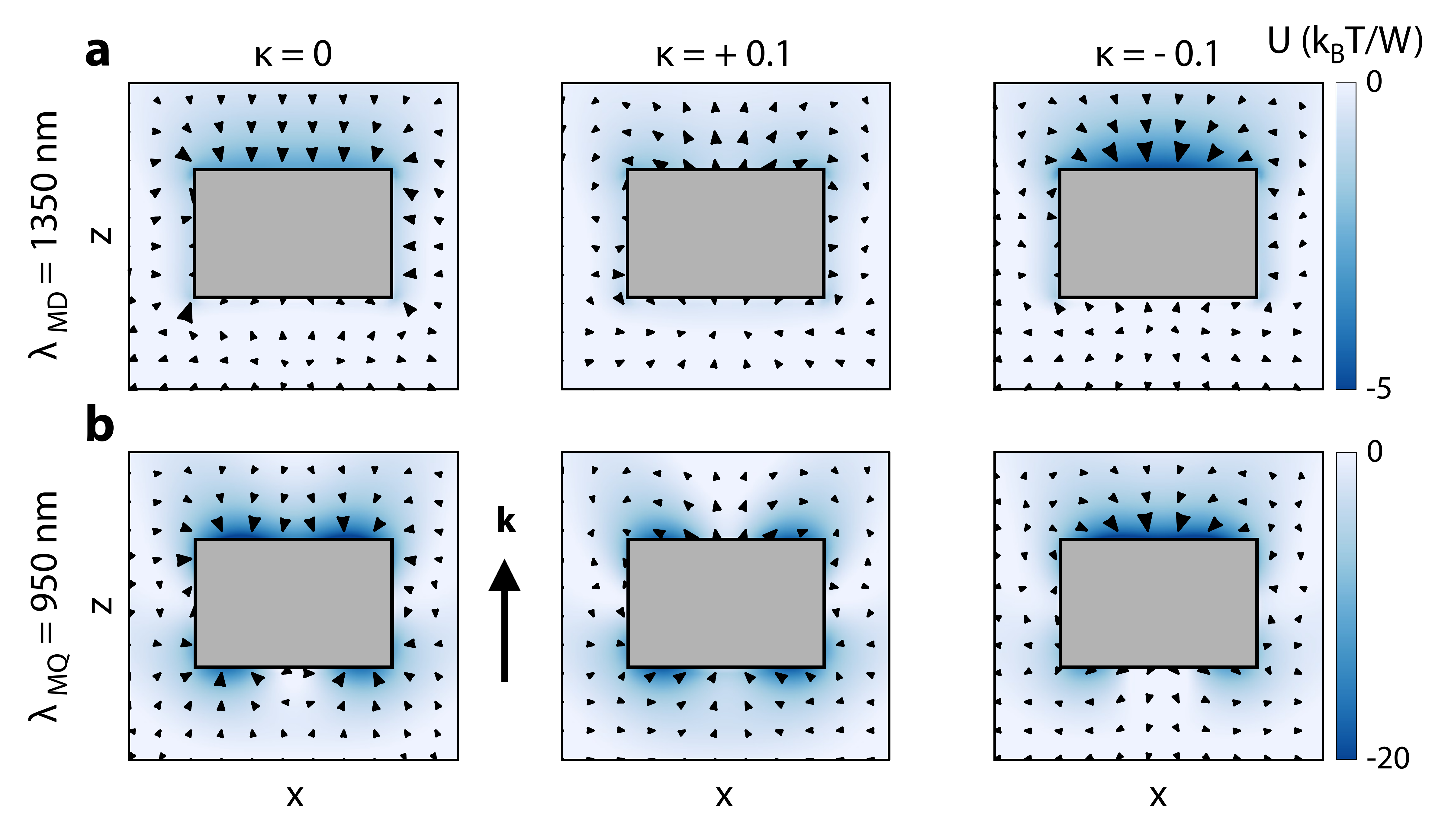}
    \caption{Force and potential on 30 nm radius PS spheres in the vicinity of the nanodisks. a) Achiral (left) and chiral (center and right) total forces and potentials for the MD resonance wavelength of 1350 nm. b) Achiral (left) and chiral (center and right) total forces and potentials for the MQ resonance wavelength of 950 nm.}
    \label{fig:2}
\end{figure*}

To quantify how these field distributions translate into mechanical action on nanoscale analytes, we evaluated the optical forces and the corresponding potentials experienced by small chiral spheres. Figure \ref{fig:2} shows the force vector fields (black arrows) acting on a polystyrene (PS, $n_{PS} = 1.57$), achiral ($\kappa = 0$) sphere of radius 30 nm; and compared it against chiral spheres characterized with $\kappa = \pm0.1$. Assuming the reactive regime, i.e., negligible dissipative and spin terms in equation \ref{eq:3}, (see Supporting Information, appendix C, for details), the total force is integrated to obtain a scalar potential $U$, which is depicted in the blue colorscale.

As shown in Figure \ref{fig:2}a, for the MD resonance, the forces mostly point vertically towards the upper face, creating a shallow potential around $-3k_B T$. This landscape changes markedly once the chiral contributions are included. In particular, the forces for the positive $\kappa$ enantiomer gain higher horizontal components, directing the particle to the outer part of the disk. On the other hand, the negative $\kappa$ enantiomer faces stronger vertical forces against the upper face of the disk, with these forces also being directed horizontally towards the center. This creates an attractive potential well around $-5k_B T$.

For the MQ resonance, in the panels below in Figure \ref{fig:2}b, forces also push particles towards the upper face of the disk. However, the intense hotspots out of the center of the disk create substantially deeper potentials, something that becomes particularly evident for the $\kappa = +0.1$ enantiomer. For the $\kappa = -0.1$ enantiomer, however, the strong central OCD gradient in Figure \ref{fig:1}f produces an attractive force, creating a deeper potential around $-25 k_B T$. These results indicate that MQ resonances are particularly well-suited for enantioselective optical trapping.

\section{Resonance tuning for optimal chiral trapping}

To find the optimal configuration for enantioselective optical trapping, we performed FEM simulations (COMSOL Multiphysics) sweeping the silicon nanodisk height $h$. To emulate realistic experimental conditions, we introduce a glass substrate which surrounds the whole silicon disk except for the upper face, where analytes are immersed in an aqueous medium. We note that the embedded fabrication of these structures can be achieved with multiple-step lithographic procedures, where electron beam lithography is combined with atomic layer deposition and other fabrication techniques to achieve great control over the glass embedding \cite{Subramania2004, Pfeiffer2016, Tanaka2020}. Furthermore, the low refractive index contrast between glass and water implies low deviations of the silicon disk response with respect to their with respect to their response in a homogeneous environment in Figures \ref{fig:1} and \ref{fig:2}.

\begin{figure*}
    \includegraphics[width=\textwidth]{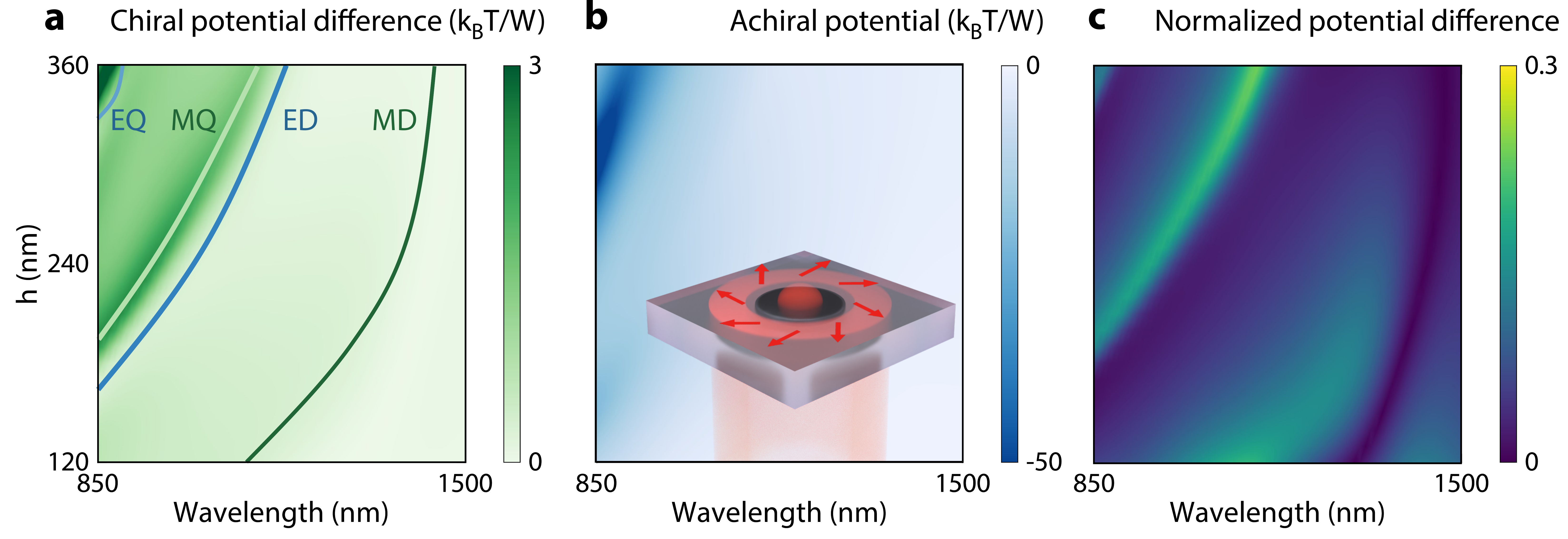}
    \caption{Evolution of the chiral and achiral potentials with disk height. a) Chiral potential difference $\Delta U_\kappa = \left| U_+ - U_- \right|$, with the evolution of resonance’s spectral positions represented by the solid lines. The Pasteur parameter was set to $\kappa = \pm 0.01$. b) Achiral potential term $U_0$. c) Normalized potential difference $\Delta U_\kappa /U_0$.}
    \label{fig:3}
\end{figure*}

To identify the optimal configuration for enantioselective trapping, we calculated the potentials from the chiral and achiral force terms in equation \ref{eq:3} at the center of the disks’ upper faces for each simulation. To be consistent with the previous section, we modeled PS nanospheres with a radius of 30 nm and a Pasteur parameter $\kappa = \pm 0.01$. The chiral potential difference $\Delta U_\kappa = \left|U_+ - U_- \right|$ is shown in Figure \ref{fig:3}a, with solid lines tracking the spectral evolution of the different multipolar maxima. While strong chirality forces are found close to the magnetic dipole resonance, as depicted in Figure \ref{fig:2}, the MQ resonance exhibits a significantly stronger potential difference. This confirms that higher-order magnetic modes are superior for maximizing chiral light-matter interactions in this geometry. 

We then compare this with the achiral potential term $U_0$ (Figure \ref{fig:3}b), which for high powers presents a relatively shallow landscape, only overcoming the trapping threshold close to the quadrupole resonances, especially the electric quadrupole (EQ) resonance. To find an optimum configuration, we plot in Figure \ref{fig:3}c the normalized difference $\Delta U_\kappa /U_0$, where the magnetic resonances are clearly emphasized. Notably, the MQ resonance displays a relatively high normalized difference, reaching up to 0.3 for a relatively small Pasteur parameter value at the tallest considered disks.

As explained before, despite using high power beams, the total potential achieved for these illumination conditions, corresponding to low Numerical Apertures (NAs), at the MQ resonance is quite shallow, meaning that particles are never trapped due to Brownian fluctuations at room temperature. To address this, we investigated the effect of increasing the NA of the incident ARPB, thereby focusing the electromagnetic energy more tightly onto the nanodisks. We focused on the optimized geometry ($h = 350$ nm) at the MQ resonance (1130 nm), which exhibited one of the highest normalized chiral potential differences; and explored NAs from 0.3 (corresponding to $w_0 = \lambda$) to 0.64 (corresponding to $w_0 = \lambda/2$).

\begin{figure*}
    \includegraphics[width=\textwidth]{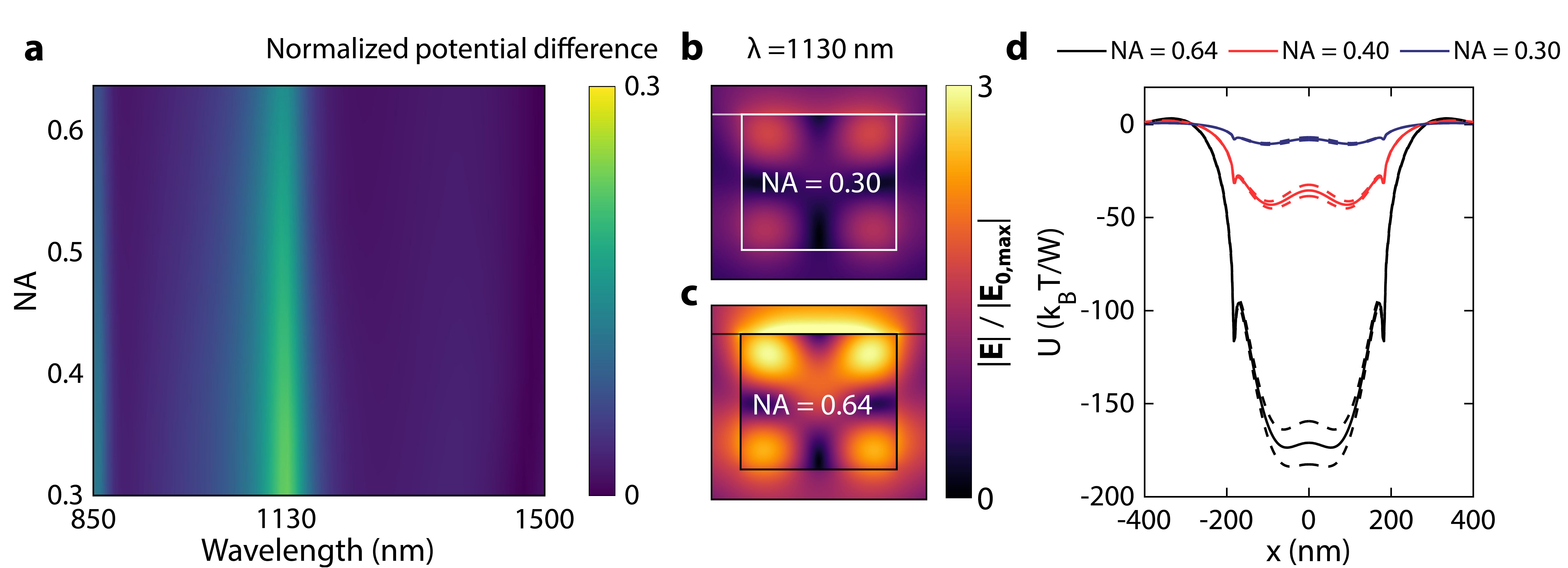}
    \caption{Evolution of the potentials with beam focusing in nanodisks with height $h$ = 350 nm. a) Normalized potential difference $\Delta U_\kappa /U_0$. b) Near electric field magnitude at the weakest ($\textrm{NA} = 0.3$) and c) strongest ($\textrm{NA} = 0.64$) focusing strengths considered in this work. d) Transversal profiles of the potentials at different focusing strengths. The solid lines represent the achiral potentials, and the dashed lines depict the chiral modifications assuming $\kappa = \pm 0.01$.}
    \label{fig:4}
\end{figure*}

The dependence of the normalized chiral potential difference on the focusing conditions is illustrated in Figure \ref{fig:4}a. We observe that the MQ resonance, fixed at 1130 nm, does not vary dramatically with the focusing strength. As NA increases, the normalized chiral difference decreases slightly—from $\sim 0.3$ at $\textrm{NA} = 0.3$ to $\sim 0.15$ at $\textrm{NA} = 0.64$. This reduction is attributed to the spatial redistribution of the electric field energy. As evidenced by the near-field maps in Figures \ref{fig:4}b-c, tighter focusing significantly amplifies the electric field intensity, concentrating it at the center of the disk's upper face. Since the achiral gradient force scales with $\nabla |\mathbf{E}|^2$, this intensity increment boosts the achiral background force relative to the chiral gradient one. The electric field intensity within the disk, however, remains consistent across all NAs, and thus confirms that the excitation of the longitudinal MQ resonance still occurs.

The slight loss in relative selectivity is compensated by a strong enhancement in trap sharpness and depth. As depicted in the transverse potential profiles (Figure \ref{fig:4}d), the potential landscape undergoes a significant shape transformation: the shallow, double-well profile characteristic of low-NA excitation, from which particles may escape easily, progressively evolves into a very deep, robust single-well trap at high NA. These deeper potentials from higher NAs enable stable trapping even for very small nanoparticles, preventing the need for excessive illumination power.

\section{Enantioseparation dynamics in the optical trap}

For realistic chiral analytes with modest Pasteur parameters ($\kappa \ll 1$), the chiral potential difference is often small compared with the achiral contribution, even in optimal conditions. Consequently, if the laser power is tuned to barely overcome the trapping threshold ($U \sim 10 k_B T$), the chiral potential difference cannot reliably provide instantaneous enantioseparation. Instead, the potential difference will asymmetrically impact the mean escape time of enantiomers in the trap. To realistically account for this effect, we employ Kramers’ reaction-rate theory to calculate the mean residence time of a particle in the trap. In the limit of strong friction (overdamped limit), the mean escape time $\tau_{esc}$ from a potencial well (position $x_0$) is given by \cite{Sicard2018}

\begin{equation}
\label{eq:4}
    \tau_{esc} = \frac{2\pi\gamma}{\sqrt{U^{''}(x_0) \left|U^{''}(x_{esc})\right|}}e^{-U(x_0)/(k_BT)}
\end{equation}
where $\gamma$ is the damping constant of the particle; and $U^{''}(x_0)$ and $U^{''} (x_{esc})$ are the potential curvatures at the potential well and the escape point, respectively. Due to the difference in potential between the two enantiomers, two different escape times $\tau_{esc}^-$ and $\tau_{esc}^+$ will exist. This asymmetry allows us to quantify the quality of the optical trap via a selectivity

\begin{equation}
\label{eq:5}
    \Gamma = \frac{\tau_{esc}^-}{\tau_{esc}^+}
\end{equation}
where the time for the negative $\kappa$ enantiomer, expected to be larger as its potential is deeper (as seen in Figure \ref{fig:2}), is placed above the positive $\kappa$ one. To assess the feasibility of practical enantioselective trapping experiments, we calculate the selectivities for the proposed $h$ =  350 nm disk at the MQ resonance wavelength of 1130 nm, plus other two nearby wavelengths (1080 nm and 1180 nm). To enable minute-long measurements, we set a threshold of 60 s for both $\tau_{esc}^+$ and $\tau_{esc}^-$. This ensures that we filter out regimes where the trap is too weak or strong to hold any particle, focusing solely on the capability to discriminate between the trapped particles and the escaping ones. Naturally, this sets a lower bound of 1 to the selectivity. A $|\kappa|=[10^{-3},1]$ range is explored, which is representative of weakly-to-moderately chiral nanoparticles, including chiral quantum dots, carbon nanotubes, NaClO\textsubscript{3} crystals, and nanoparticles functionalized with chiral ligands or assembled with DNA techniques \cite{Zafar2020,Varga2017,Wei2016,Niinomi2021}. These systems occupy an intermediate regime between naturally occurring molecules, with much smaller Pasteur parameters, and highly engineered chiral nanostructures, whose larger chiral responses are typically linked to resonant features, and therefore do not fit the assumptions made in equation \ref{eq:3}. Furthermore, for the practical demonstration of our approach, and in accordance with standard considerations in other relevant literature \cite{Martnez-Romeu2024, Zhao2016, Xiong2024, Donato2014, Fang2021, Martinez-Romeu2026, Cheng2024}, the beam power is lowered to 100 mW.

\begin{figure*}
    \includegraphics[width=\textwidth]{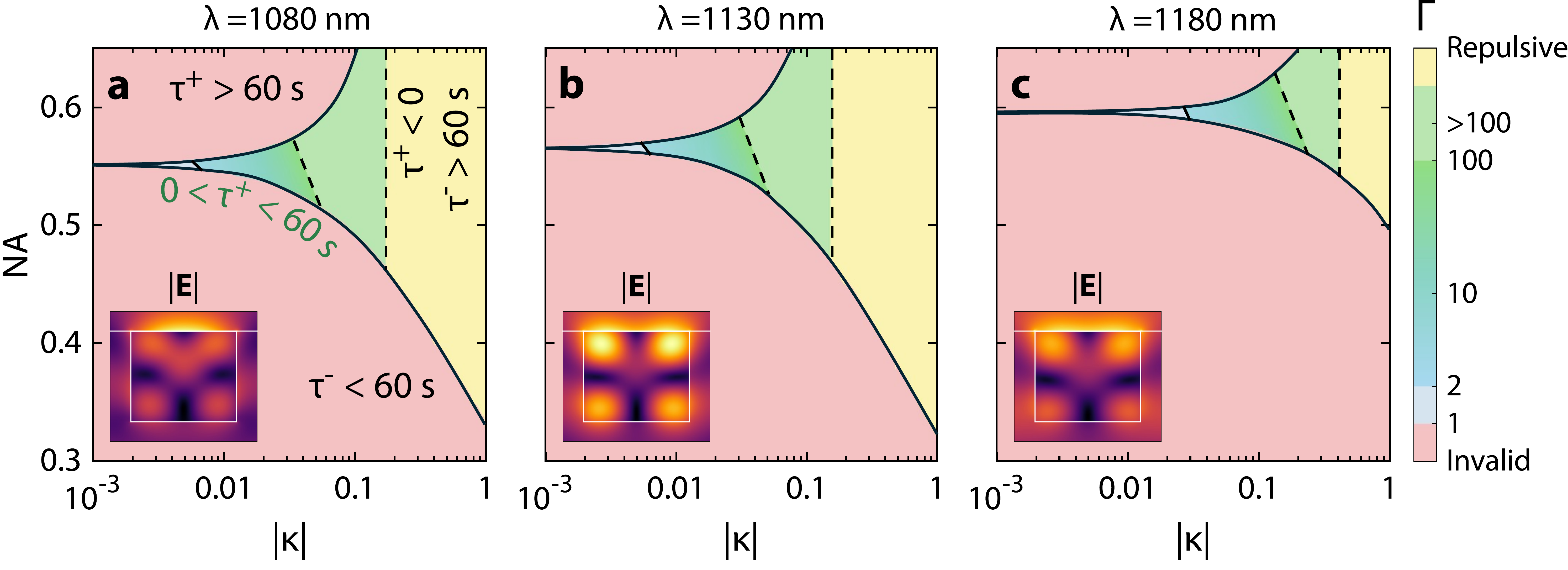}
    \caption{Trap selectivity for different illumination (NA) and target ($|\kappa|$) conditions, at a) 1080 nm, b) 1130 nm and c) 1180 nm wavelengths. The red regions represent points of the parameter space where enantioselective trapping is not possible, either because the mean escape time for the undesired enantiomer is longer than 60 s, or the mean escape time for the desired enantiomer is shorter than 60 s. On the other hand, other colors represent points where the two conditions are met, allowing enantioselective trapping: the blueish gray areas correspond to inefficient selectivities between 1 and 2, while the blue to green gradient represents selectivities from 2 to 100 in a logarithmic scale. A light green color is employed for selectivities higher than 100 and the yellow region corresponds to negative mean escape times for the undesired enantiomer, meaning that the undesired enantiomer is repelled and the selectivity is virtually infinite. The insets depict the electric near field magnitude at the corresponding illumination wavelengths.}
    \label{fig:5}
\end{figure*}

The calculated selectivities are shown in Figure \ref{fig:5}. For all wavelengths, the parameter space differentiates between several different regions. The instability regime (red regions) corresponds to conditions where the optical potential is: a) insufficient to retain the target enantiomer against thermal fluctuations for the required 60 s threshold; or b) excessive, meaning that both enantiomers are trapped for times over the 60 s threshold. 

A second region, corresponding to an intermediate regime, appears as a blue-to-green gradient. In this regime, the conditions for selective trapping are met, and the selectivity ranges from 1 to 100. For practical purposes, this region is separated into three different areas: one (blueish gray) where the selectivity is between 1 and 2, and therefore selective trapping is not reliable; another region with the 2 to 100 range, displayed in a logarithmic scale; and another (light green) where selectivities are above 100, and therefore selective trapping is excellent. Finally, a third region, in yellow, corresponds to situations where the undesired enantiomer is repelled rather than trapped, resulting in virtually infinite selectivities.

As seen in Figure \ref{fig:5}a-b, at both resonant wavelengths (1080 nm and 1130 nm), enantioselective trapping becomes possible for Pasteur parameters as low as $|\kappa| < 0.01$. In particular, at 1130 nm, the theoretical detection limit is around $|\kappa|= 0.001$, with selectivity $\Gamma = 1$. For realistic trapping, requiring a higher $\Gamma \geq 2$ value, $|\kappa|$ must be slightly increased to $|\kappa| \simeq 0.006$. More stable trapping might be found at higher values of $|\kappa|$, like $\Gamma = 10$ at $|\kappa| \simeq 0.02$. These values indicate that the platform is sensitive beyond artificial chiral media. Beyond these thresholds for selective trapping, both wavelengths show excellent selectivities (above 100) at $|\kappa| \simeq 0.03-0.04$; and feature a transition to the infinite regime at $|\kappa| \simeq 0.15$.

Another interesting feature is that, as shown by the insets, the electromagnetic field becomes more tightly focused on the upper face center as the wavelength is reduced, resulting in deeper potentials. This allows for the use of lower Numerical Apertures without compromising trap stability for the same input power. Conversely, at a particular wavelength, more power can be employed to reach the same selectivity levels with a lower NA (see Supporting Information, appendix E, for details). On the other hand, the higher wavelength of 1180 nm in Figure \ref{fig:5}c shows the opposite effect: the electric focusing is weaker and higher NAs are required. Due to this, and in addition to being out of resonance, the chiral-to-achiral force ratio is lower and poorer selectivities are featured.

In addition to the enhanced enantioselectivity of this approach, our all-dielectric platform lowers the risk of photodisruption or thermal damage \cite{Matsuura2015}, as opposed to conventional plasmonic tweezers under high optical powers \cite{Lin2021,Zhao2016}. As shown in the Supporting Information (appendix F), the negligible absorption coefficient of silicon in the near-infrared regime prevents significant temperature increments, even under high power illumination. This ensures that, even under the high powers required for low NAs, high selectivity trapping can be safely achieved without compromising the structural integrity of delicate samples.

\section{Conclusion}

We have demonstrated that silicon nanodisks illuminated by Azimuthally–Radially Polarized Beams provide a robust platform for enhanced enantioselective optical manipulation. By exciting longitudinal Mie resonances, specifically the Magnetic Quadrupole (MQ) mode, we engineered optical fields that exhibit intense gradients of optical chirality while maintaining comparatively moderate electric-field intensity gradients. Through a systematic optimization of the nanodisk geometry, beam numerical aperture, and illumination wavelength, we identified optimal trapping conditions where the chiral potential differences become significant, even for analytes with modest Pasteur parameter values.

Furthermore, by integrating Kramers’ escape-rate theory, we have further demonstrated the practical implementation of our design. Our results predict that under experimentally realistic powers, high enantioselectivities (over 100) are feasible for $|\kappa| \geq 0.03$. Moreover, reliable discrimination (above 2) remains achievable for $|\kappa|$ in the $0.005-0.01$ range. Importantly, the low-loss nature of silicon ensures minimal photothermal heating, thereby preserving the structural integrity of delicate samples, a significant advantage over plasmonic alternatives for biological and chemical applications. We note that, although the operating optical power is, like most optical approaches, relatively high (100 mW), the usage of higher NAs could significantly lower this requirement. Although higher NAs show a deterioration of the chiral-to-achiral force ratio, robust trapping should still be achieved in the $|\kappa| \sim 10^{-2}$ range.

Another advantage lies in the spatial positioning of the trapping gradients, as longitudinal resonances place these on the upper faces of disks, unlike previous proposals that rely on the lightning rod effect and thus require a very precise positioning of the analytes \cite{Yamane2023}. In this respect, this approach could be practically implemented in an optical sorting scheme with microfluidic channels \cite{Donato2014,Tkachenko2014}. To account for the low throughput of a single disk, an array of disks could be fabricated, and an independent illumination of each disk can be achieved by means of holographic techniques \cite{Jones2015}.

Overall, this work establishes longitudinal Mie resonances in high-index dielectrics as a promising strategy for non-invasive, power-efficient, and thermally stable optical enantioseparation. These findings pave the way for next-generation applications in sensing, pharmacology, and optical manipulation in micro-fluidic systems.

\section{Methods}

All numerical electromagnetic calculations in this work were performed by electrodynamic simulations in COMSOL (RF Module), providing an accurate solution of the Maxwell equations. Due to the cylindrical symmetry of the problem, a 2D axisymmetric configuration was selected. The paraxial expression described in equation \ref{eq:1} was employed to calculate the beams’ background fields in a scattered field formulation. Due to the presence of the glass-water interface, the background field was introduced with a piecewise definition that considered both the refractive index differences as well as the Fresnel reflection and transmission coefficients. Details concerning the accuracy of this method against full-field simulations are discussed in the Supporting Information (appendix G). 

A beam waist $w_0 = \lambda/2$ (corresponding to $\textrm{NA} = 0.64$) was considered as a limit for paraxial beams, as 95\% of the beam energy is carried out in propagating modes for such beams \cite{Veysi2016}. For silicon, material data from Aspnes et al. \cite{Aspnes1983} was considered, while constant refractive indices were considered for glass ($n_G= 1.5$) and water ($n_W= 1.33$). Multipole analysis was performed by calculating the internal currents of the nanodisks in a homogeneous medium following \cite{Evlyukhin2016}.

Forces were calculated using equation \ref{eq:3} while neglecting dissipative terms (as justified in the Supporting Information). The following expressions were used for the polarizabilities of a small sphere with radius $r$ \cite{Genet2022}:

\begin{subequations}
\begin{equation}
    \alpha_e = 4\pi\varepsilon_0 r^3 \frac{(\varepsilon_r - \varepsilon_m)(\mu_r + 2) - \kappa^2}{(\varepsilon_r + 2\varepsilon_m)(\mu_r + 2) - \kappa^2}
\end{equation}

\begin{equation}
    \alpha_{em} = 12\pi r^3 \frac{\kappa\sqrt{\varepsilon_0 \mu_0}}{(\varepsilon_r + 2\varepsilon_m)(\mu_r + 2) - \kappa^2}
\end{equation}
\end{subequations}
where $\varepsilon_r = n_{PS}^2$ is the relative permittivity of the PS sphere, $\varepsilon_m=n_W^2$ is the relative permittivity of the background medium (water) and $\mu_r= 1$. It should be noted that the cylindrical symmetry of the problem prevents forces from having azimuthal components. As only conservative terms remained in the calculation of forces, potentials were calculated by numerical integration of the forces $U_{\rho z} =\int F_\rho d\rho + \int F_z dz$. Integration in the radial domain was performed by considering the $\rho = 400$ nm endpoint shown in Figure \ref{fig:4}d as a zero-potential point, while integration in the $z$ direction was considered only from the upper face level to 200 nm above, which was fixed as a zero potential level as well.

To compute escape times, we fit the bistable radial potential profiles to a quartic analytical form $U(\rho)= a \rho^4 + b \rho^2 + c$ \cite{Jones2015}, from which the curvatures at the potential minima $U^{''}(x_0)$ could be easily extracted. The escape point curvatures $|U^{''}(x_{esc})|$, on the other hand, were obtained from numerical derivatives at potential maxima found beyond the disk radius. More details can be found in the Supporting Information (appendix D). The damping constant of the spherical particle was obtained as $\gamma = 6\pi \eta_W r$, with $\eta_W= 0.001 \ \textrm{Pa} \cdot \textrm{s}$ being the water viscosity \cite{Jones2015}. A constant room-temperature $T = 300$ K was considered throughout this work.

\textbf{Acknowledgements} \par 
This work acknowledges funding by the MOPHOSYS Project (PID2022-139560NB-I00) from Proyectos de Generación de Conocimiento provided by the Spanish Agencia Estatal de Investigación. G.S. thanks the Spanish Ministry of Education for his predoctoral contract grant (FPU21/02296).

\appendix
\section{Full paraxial ARPB expression}

Equation \ref{eq:1} in the main text reads

\begin{multline}
    \label{eq:s1}
    \mathbf{E^{ARPB}} = \mathbf{E^{APB}} + \mathbf{E^{RPB}} = \frac{V_R}{w^2} f(A_\rho + i B_\rho) \boldsymbol{\hat{\rho}} + \\ + \frac{V_A}{w^2}f e^{i\Psi} \boldsymbol{\hat{\varphi}} + \frac{2iV_R}{kw^2} f(A_z + iB_z) \mathbf{\hat{z}}
\end{multline}
where $V_R$ and $V_A$ are volt-unit amplitudes for the radial and azimuthal polarizations and $\Psi$ is a constant phase difference between them \cite{Herrero-Parareda2024}. $w = w_0 \sqrt{1+ (z/z_R)^2}$ is the beam radius, which at the focus point is equal to $w_0$; and $z_R = \pi w_0^2/\lambda$ is the Rayleigh range of the beam. The shorthand $A$ and $B$ parameters, plus the Gaussian envelope $f$, are given by

\begin{subequations}
\label{eq:s2}
\begin{equation}
    A_\rho = 1+ \frac{1}{kz_R}\frac{\rho^2-2w_0^2}{w^2} +\left( \frac{2z\rho}{w^2 kz_R} \right)^2
\end{equation}

\begin{equation}
    B_\rho =- \frac{4}{k^2 w^2}  \frac{z}{z_R}  \left( 1 - \frac{\rho^2}{w^2} \right)
\end{equation}

\begin{equation}
    A_z = 1 - \frac{\rho^2}{w^2}
\end{equation}

\begin{equation}
    B_z= \frac{z}{z_R} \frac{\rho^2}{w^2}
\end{equation}

\begin{equation}
    f = \frac{2}{\sqrt{\pi}} e^{-\left( \frac{\rho}{w} \right)^2 \zeta - 2i \tan^{-1} \left( \frac{z}{z_R} \right) + ikz}
\end{equation}
\end{subequations}
with $k$ being the wavenumber and $\zeta = 1-i z/z_R$  being the Gouy phase. The total power of the beam, in a medium with impedance $\eta$, is given by

\begin{multline}
    P_{ARPB} = P_{APB} + P_{RPB} = \\ = \frac{1}{2\eta} \left(|V_A|^2+ |V_R|^2 \right) \left( 1-\frac{1}{\omega z_R} \right)
\end{multline}

\section{Cartesian multipole components}
\renewcommand{\thefigure}{B\arabic{figure}}
\setcounter{figure}{0}

\begin{figure}
    \centering
    \includegraphics[width=\linewidth]{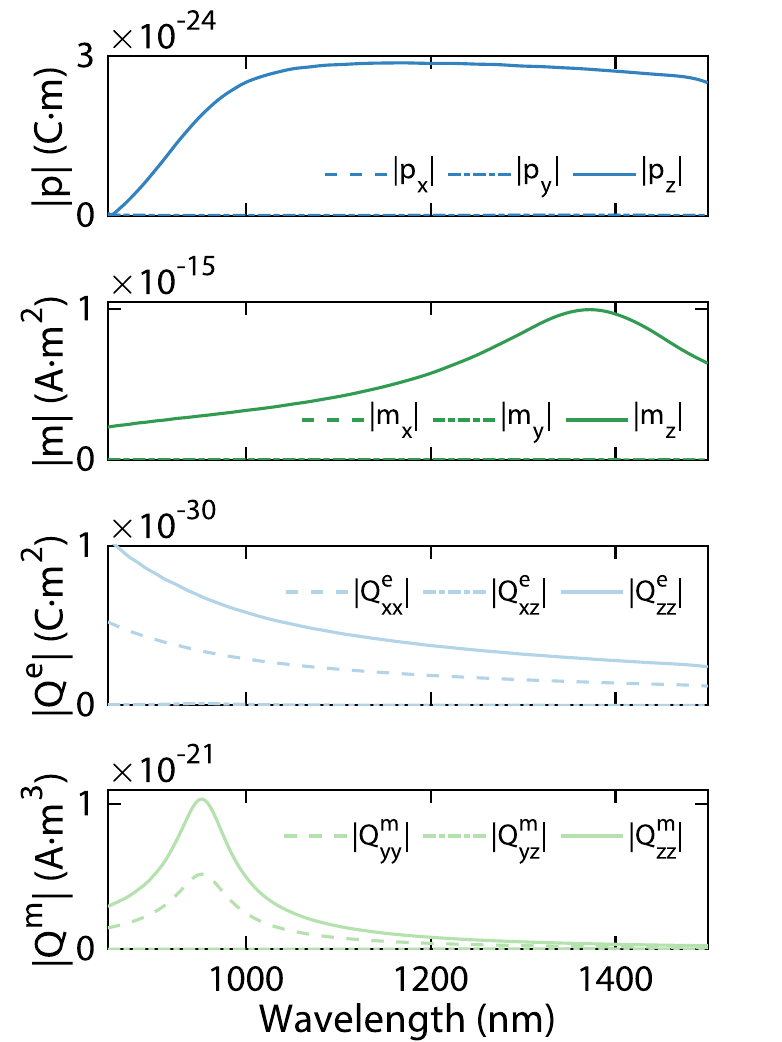}
    \caption{Absolute value of the different Cartesian components of the multipole moment tensors in Figure \ref{fig:1}c of the main document. For dipolar resonances, characterized by first-rank tensors (vectors), all three components are depicted. For quadrupolar resonances, characterized by second-rank tensors, only the three independent components (diagonal $xx/yy$, diagonal $zz$, and off-diagonal $xz/yz$) are featured.}
    \label{fig:S1}
\end{figure}

The scattered powers depicted in Figures \ref{fig:1}a, \ref{fig:1}b and \ref{fig:1}c in the main document come from the computation of the multipolar moment tensors from the internal electric fields, as shown in \cite{Evlyukhin2016}. Figure \ref{fig:S1} contains a decomposition of the cartesian components of the different multipolar moment tensors, allowing us to discern whether the excited resonances are truly of longitudinal nature.

For first-order resonances (electric and magnetic dipoles, corresponding to first-rank tensors), Figure \ref{fig:S1} contains all three elements ($x$, $y$ and $z$), with the longitudinal $z$ component being the only relevant contributions. 

On the other hand, quadrupolar resonances (corresponding to second-rank tensors) display two relevant contributions. While the longitudinal $zz$ elements are still the largest contributors, other diagonal tensor elements are also important. Off-diagonal terms, often the dominant contributors with plane wave illuminations, are negligible with ARPB illumination. It should be noted that, due to the cylindrical symmetry of the system, $|Q_{xx}^{e,m}| = |Q_{yy}^{e,m}|$ and $|Q_{xz}^{e,m}| = |Q_{zx}^{e,m}| = |Q_{zy}^{e,m}| = |Q_{yz}^{e,m}|$, so Figure \ref{fig:S1} shows all independent components of the tensors.

\section{Dissipative and spin density force contributions}
\renewcommand{\thefigure}{C\arabic{figure}}
\setcounter{figure}{0}

As noted in the main document, the force in its equation \ref{eq:3} can be separated into conservative reactive forces, depending on the electric field intensity and optical chirality gradients:

\begin{multline}
    \langle \mathbf{F} \rangle_{reac} = \frac{1}{4} \mathfrak{Re}(\alpha_e) \nabla |\mathbf{E}|^2 - \\ - \frac{1}{2} \mathfrak{Re}(\alpha_{em}) \nabla \left[\mathfrak{Im}(\mathbf{E^*} \cdot \mathbf{H})\right] 
\end{multline}
and dissipative forces, coming from scattering and the electric field spin-density. 

\begin{equation}
    \langle \mathbf{F} \rangle_{diss} = \sigma \frac{\langle \mathbf{S} \rangle}{c} + \omega \gamma_e \langle \mathbf{L_e} \rangle 
\end{equation}

\begin{figure*}[!t]
    \centering
    \includegraphics[width=\textwidth]{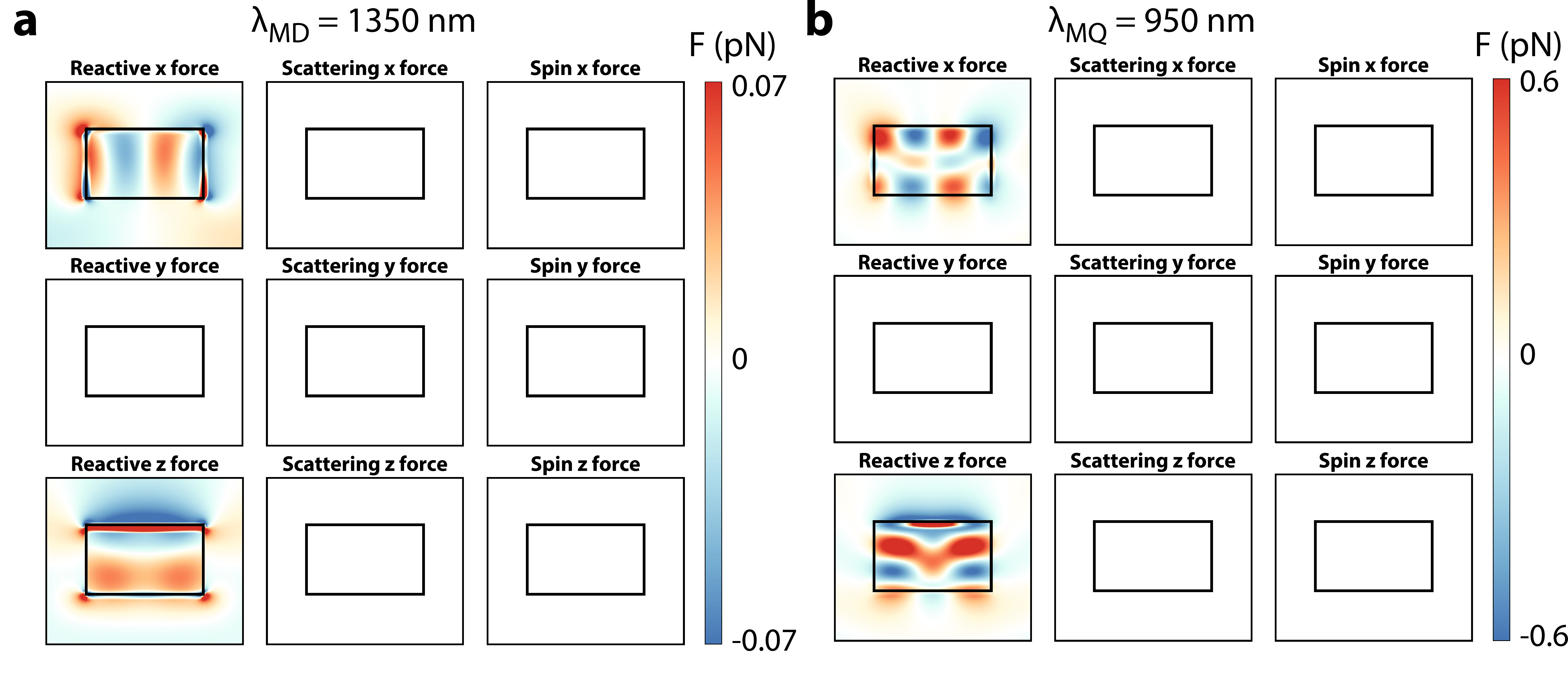}
    \caption{Decomposition of the forces displayed in Figure \ref{fig:2} of the main document in their cartesian components and reactive and dissipative (scattering and spin) terms.}
    \label{fig:S2}
\end{figure*}

Integration of forces to compute a potential is only possible if these dissipative forces are negligible against the reactive ones. To check that this condition is met, a decomposition of these force terms is shown in Figure \ref{fig:S2} for the two wavelengths analyzed in Figures \ref{fig:1} and \ref{fig:2} of the main document. For the magnetic dipole resonance wavelength (1350 nm), it is clear that the $x$ component of the reactive force is restorative, pushing objects to the center of the disk’s upper face, while the stronger $z$ component pushes the object towards the upper face. This is also observed at the magnetic quadrupole resonance wavelength (950 nm), where the reactive $x$ component has two equilibrium points, corresponding to the potential wells observed in Figure \ref{fig:2}. For both wavelengths, the reactive $y$ component (evaluated at the $xz$ plane) is exactly zero due to the cylindrical symmetry of the system.

More importantly, the scattering and spin dissipative terms of the force are in all cases negligible in comparison with the reactive terms, enabling us to compute potentials from the reactive terms. In particular, while reactive forces are often in the 0.01-0.1 pN range, scattering and spin forces are around five orders of magnitude smaller.

\section{Details of potential fitting}
\renewcommand{\thefigure}{D\arabic{figure}}
\setcounter{figure}{0}

\begin{figure*}
    \centering
    \includegraphics[width=0.8\textwidth]{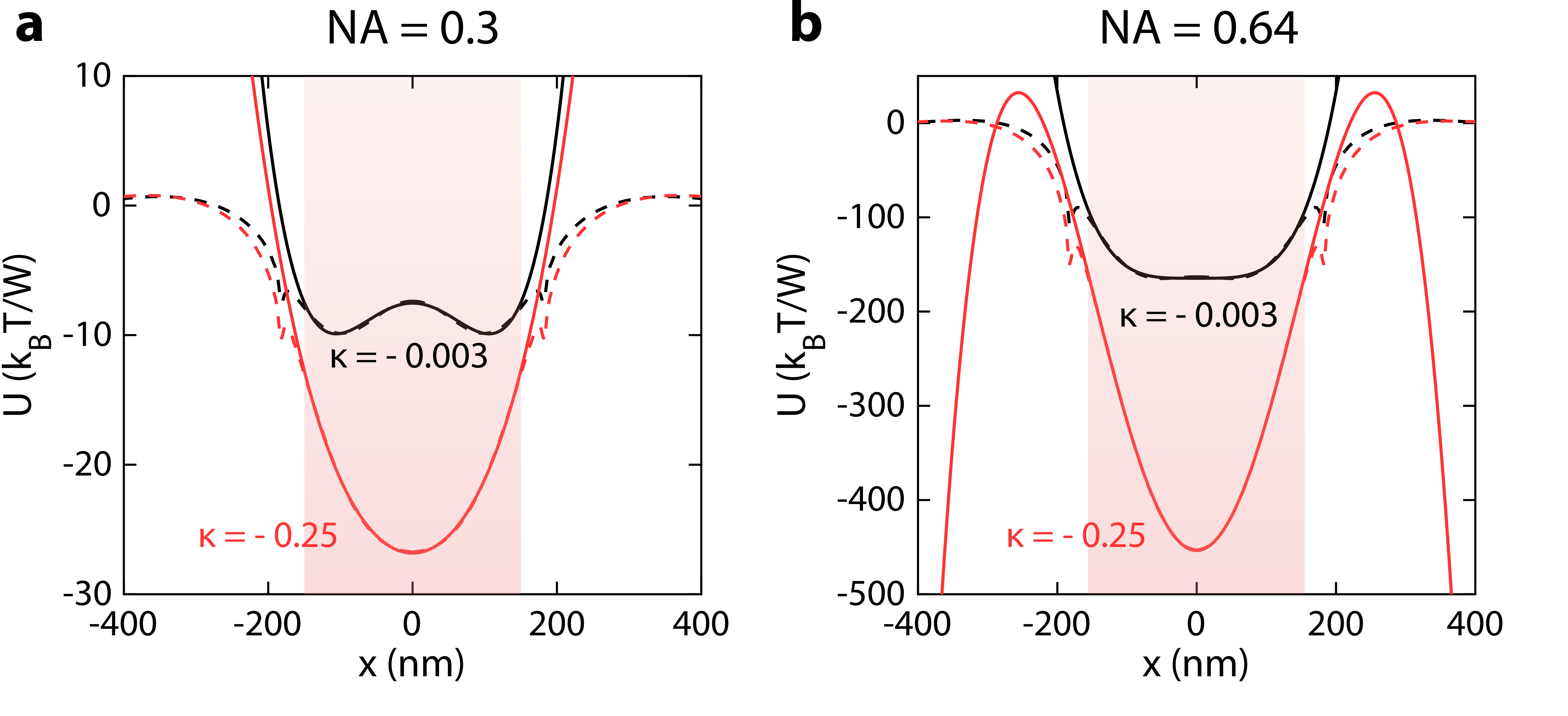}
    \caption{In dashed lines, radial profile of some of the potentials at the disk upper face, with the quartic fitting in solid lines. The fitting is only performed in the $[-150, 150]$ nm interval, represented by the reddish region in the background. a) Potential with $NA = 0.3$ illumination ($\kappa = -0.003$ in black, $\kappa = -0.25$ in red). b) Potential with $NA = 0.64$ illumination ($\kappa = -0.003$ in black, $\kappa = -0.25$ in red).}
    \label{fig:S3}
\end{figure*}

As stated in the methods section of the main document, potentials were calculated by numerical integration of the forces $U_{\rho z} =\int F_\rho d\rho + \int F_z dz$. Integration in the radial domain was performed by considering the $\rho = 400$ nm endpoint shown in Figure \ref{fig:4}d as a zero potential point, while integration in the $z$ direction was considered only from the upper face level to 200 nm above, which was fixed as a zero potential level as well. To compute escape times from the potential curvatures, we fit the bistable radial potential profiles to a quartic analytical form $U(\rho)= a \rho^4 + b \rho^2 + c$ in the $[0,150]$ nm radial interval \cite{Jones2015}. This interval allows to capture most of the disk’s upper face, while leaving outside the sharp potential spikes, which correspond to lightning rod effects. As shown for exemplary cases in Figure \ref{fig:S3}, the fit within this interval is very good, and allows to reproduce potential minima even at the weakest focusing ($NA = 0.3$), where the bistable minima are well resolved.

Such a fit allows to easily detect the potential minima and extract their curvature $U^{''}(x_0)$ could be easily extracted. For strong positive values of $\kappa$, the potential might not have a minimum in this range. This situation corresponds to a repulsion, so if no minima could be found, the curvature was not computed and instead the mean escape time was set to zero and thus an infinite selectivity. 

On the other hand, the escape point curvatures $|U^{''}(x_{esc})|$ were obtained from numerical derivatives:

\begin{equation}
    U^{''}(x_{esc})  \approx \frac{U(x_{esc} + \Delta x) + U(x_{esc} - \Delta x) - 2U(x_{esc})}{\Delta x^2}
\end{equation}

The escape points were determined by finding the potential maxima beyond the disk radius ($\rho \geq 200$ nm). In this case, a repulsion situation results in no maxima found in this range. Likewise, the curvature was not computed in those cases, resulting again in a zero mean escape time and an infinite selectivity.

\section{Optical trap dynamics for higher power illuminations}
\renewcommand{\thefigure}{E\arabic{figure}}
\setcounter{figure}{0}

The results shown in Figure \ref{fig:5} of the main document correspond to an illumination power of 100 mW, consistent with literature on enantioselective optical trapping. As noted there, the reddish region below the $|\kappa|$ minimum peak corresponds to situations where enantioselective optical trapping would be possible, but the trapping time would be insufficient (below 60 s). Such a limitation can be addressed with higher optical power, enlarging the achiral trapping potential and thus increasing the trapping time.

\begin{figure*}
    \centering
    \includegraphics[width=\textwidth]{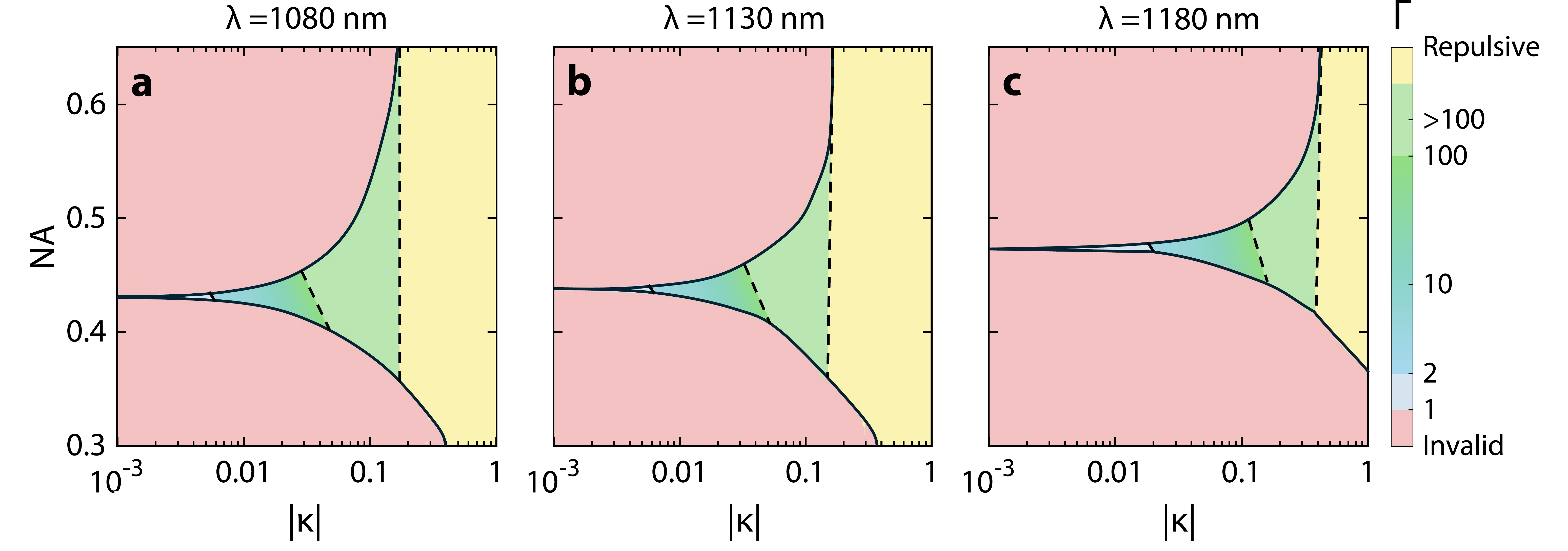}
    \caption{Trap selectivities, as depicted in Figure \ref{fig:5} of the main document, for a) 1080 nm, b) 1130 nm and c) 1180 nm. The illumination power here is set to 300 mW.}
    \label{fig:S4}
\end{figure*}

To assess the performance of our system under such conditions, we plot in Figure \ref{fig:S4} a replica of Figure \ref{fig:5} with an illumination power of 300 mW. Since the mean escape time for the undesired enantiomer also grows, the whole profile of the selectivity shifts towards lower NA values. This could be beneficial since the normalized chiral difference $\Delta U_\kappa / U_0$ was larger at smaller NAs. However, operation at such powers should be considered with caution, since the strength of the fields might cause photothermal or photodisruptive damage to the samples.

\section{Heat generation in the nanodisks}
\renewcommand{\thefigure}{F\arabic{figure}}
\setcounter{figure}{0}

Despite the low-loss nature of high refractive index dielectric materials such as silicon, the very high powers employed for optical trapping result in non-negligible temperature increments. To explore the thermal behavior of the system, stationary heat transfer simulations were carried out in COMSOL, using the absorbed electromagnetic power (resistive losses) in the silicon disks as the heat source. For silicon, a thermal conductivity of 148 W/(m·K) was considered, while 0.6 W/(m·K) and 1.4 W/(m·K) conductivities were considered for water and glass, respectively. An Interfacial Thermal Conductance (ITC) of 143 MW/(m\textsuperscript{2}·K) was considered between water and silicon \cite{Gonalves2022}.

\begin{figure*}
    \centering
    \includegraphics[width=\textwidth]{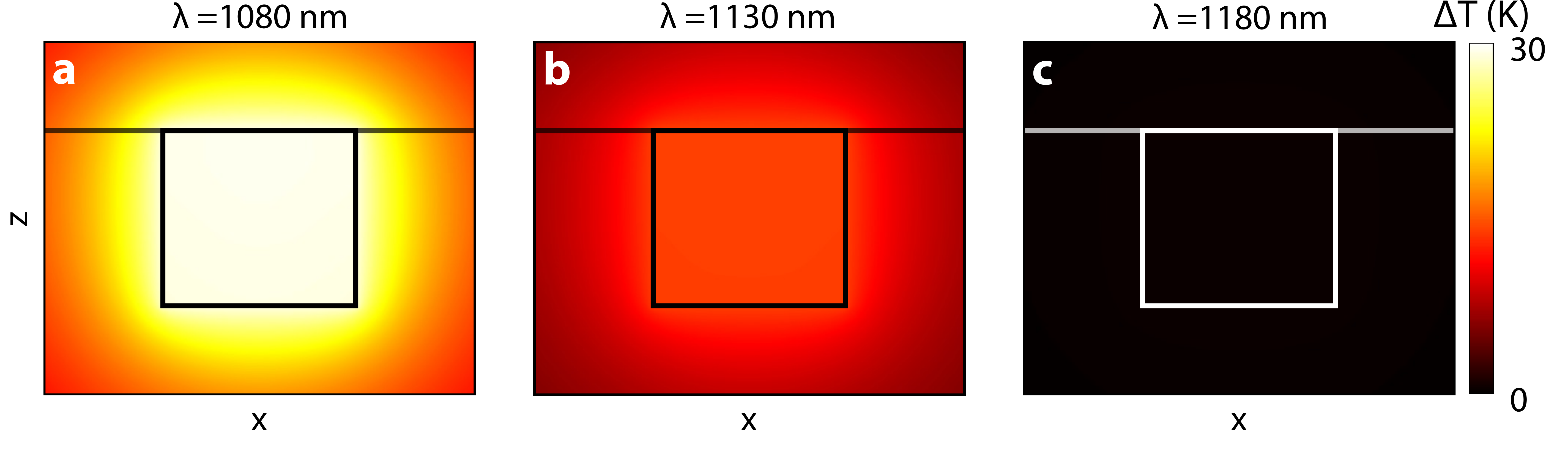}
    \caption{Temperature increment profiles for a) 1080 nm, b) 1130 nm and c) 1180 nm; at 100 mW illumination power and the maximum focusing strength considered ($NA = 0.64$).}
    \label{fig:S5}
\end{figure*}

Figure \ref{fig:S5} shows the temperature profiles for the maximum focusing strength ($NA = 0.64$) at the cases examined in Figure \ref{fig:5} of the main document. In general, temperature increments are mild despite the relatively high power and tight focusing, as silicon losses are low in this spectral range. The temperature increase at the resonant wavelength (1130 nm) reaches around 15 K, while using lower wavelengths (1080 nm) can double the heating. Conversely, off-resonant wavelengths (1180 nm) result in negligible heating. It should be noted that temperature increments of $\sim 20$ K might result in permanent damage to some compounds (for example, denaturalization in proteins \cite{Matsuura2015}), and thus, usage of larger disks, resonant at longer wavelengths where silicon losses are negligible, would be advisable. Another alternative would be to use dielectric materials with lower losses, such as GaP \cite{Aspnes1983}.

\section{Details on scattered-field simulations and comparison to full-field simulations}
\renewcommand{\thefigure}{G\arabic{figure}}
\setcounter{figure}{0}

\begin{figure*}[!t]
    \centering
    \includegraphics[width=0.5\textwidth]{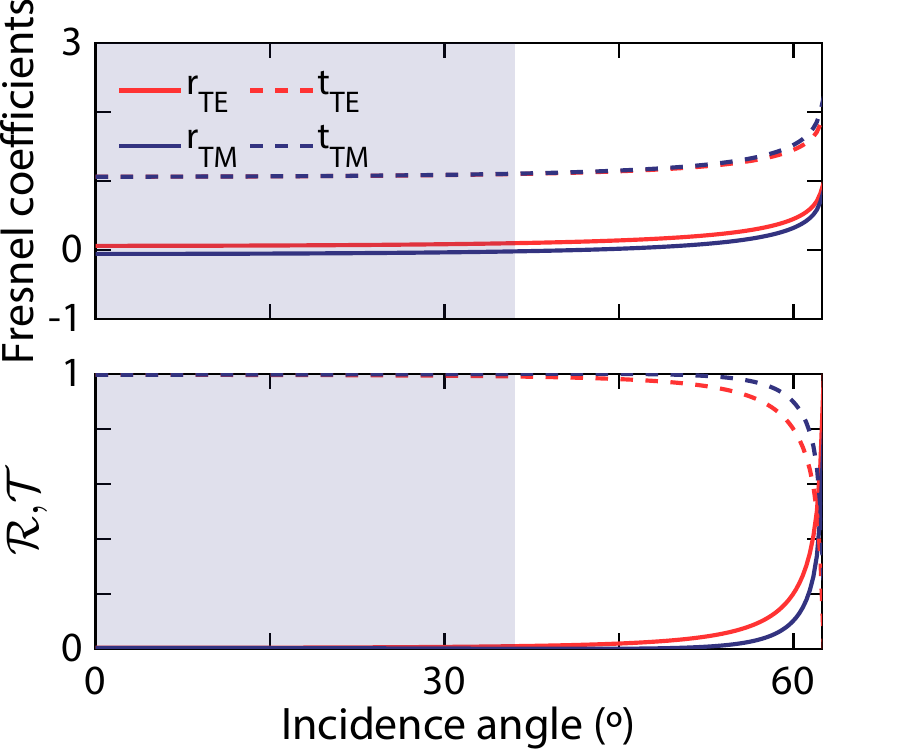}
    \caption{Fresnel coefficients (upper panel), as well as the reflection $\mathcal{R}$ and transmission $\mathcal{T}$ factors (lower panel) for both TE and TM polarizations. The shadowed area corresponds to the paraxial regime of the considered beams. For visualization clarity, only angles up to the critical angle $\theta_c = n_W/n_G \approx 62$º are displayed.}
    \label{fig:S6}
\end{figure*}

As stated in the methods section, numerical electromagnetic calculations in this work were performed in a scattered-field formulation in COMSOL (RF Module). To accurately solve this problem in the presence of a substrate, we introduce the background fields from equations \ref{eq:s1} and \ref{eq:s2} in a piecewise manner, following the constant refractive indices of glass ($n_G= 1.5$) and water ($n_W= 1.33$). To account for reflections and tranmissions, the piecewise fields were corrected by their corresponding Fresnel coefficients, shown in Figure \ref{fig:S6}. As the beam is emitted from the substrate side, the Fresnel coefficients consider glass as the first medium and water as the second.

\begin{figure*}
    \centering
    \includegraphics[width=0.9\textwidth]{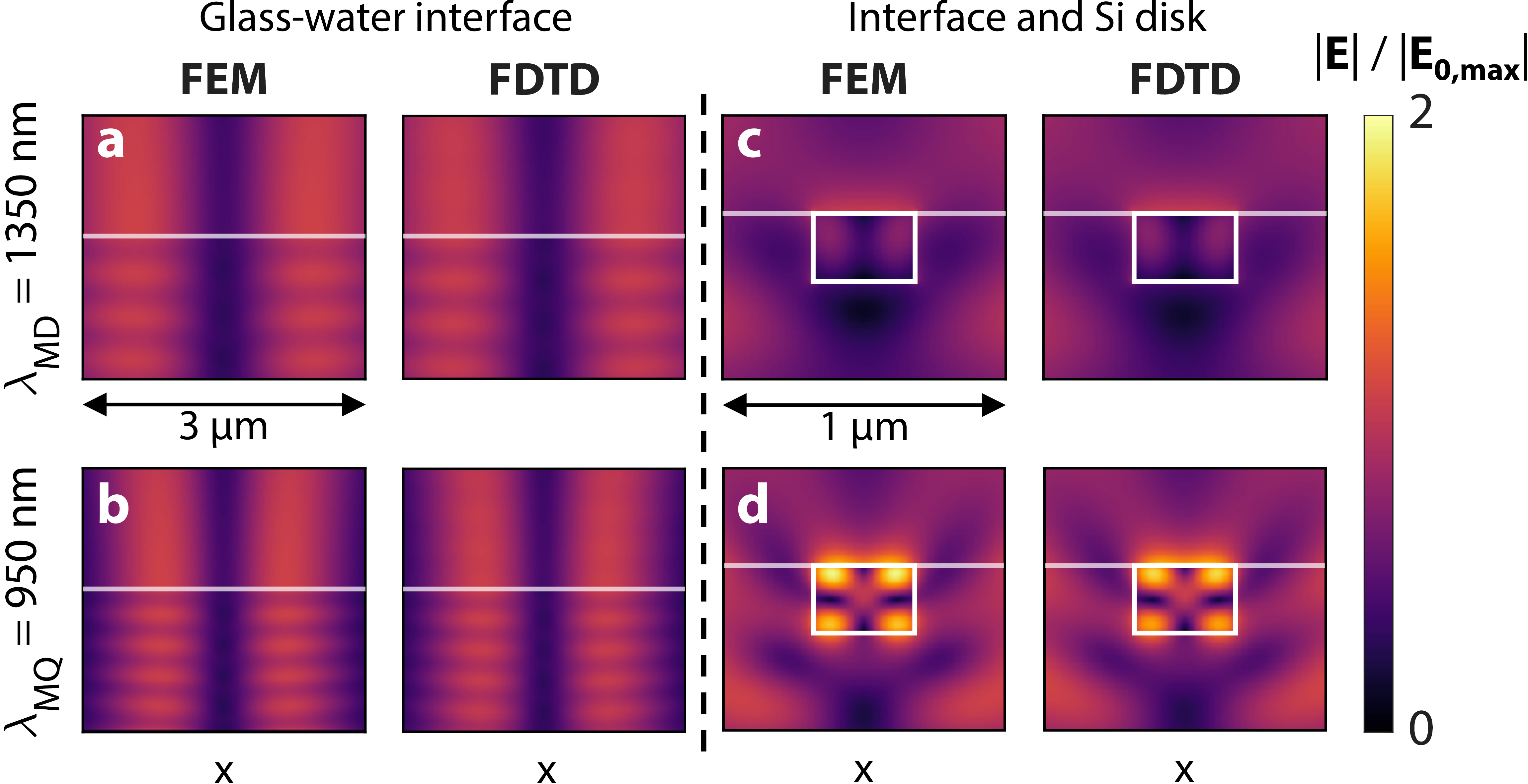}
    \caption{Transversal profiles ($XZ$ planes) of the electric field magnitude obtained by our scattered-field approach in FEM simulations (COMSOL) and a full-field approach in FDTD (Lumerical). a) Simulations considering only the glass-water interface at the magnetic dipole (a) and magnetic quadrupole (b) resonance wavelengths in Figure \ref{fig:1}. c) Simulations considering the glass-water interface plus the $h = 240$ nm silicon nanodisk at the magnetic dipole (c) and magnetic quadrupole (d) resonance wavelengths in Figure \ref{fig:1}. Panels a and b display larger regions ($3\times3$ $\mu$m images) to show the general profile of the beam, while panels c and d zoom in on the disk ($1\times1$ $\mu$m images) for a better comparison of near-field features.}
    \label{fig:S7}
\end{figure*}

As shown in this image, the coefficients assume relatively constant values across the paraxial regime ($r_{TE} \approx 0.06$, $r_{TM} \approx -0.06$ and $t \approx 1.06$ for both polarizations). Thus, the corrections applied to the beams corresponded to these constant values for all simulations. It should be noted that the APB and RPB configurations correspond to TE and TM modes, and thus the TE coefficients were assigned to the azimuthal component of the field and the TM coefficients to the radial and longitudinal components.

To check the validity of our approach, we provide a comparison with full-field simulations in Figure \ref{fig:S7}. These full-field simulations were performed with Lumerical FDTD, employing its ``Import source'', where a custom field is injected at a single plane. We check two different situations: one with only the glass-water interface (Figure \ref{fig:S7}a-b) and another with both the interface and the $h = 240$ nm disk examined in Figures \ref{fig:1}-\ref{fig:2} of the main text (Figure \ref{fig:S7}c-d). 

As shown in Figure \ref{fig:S7}, the agreement between the two methods in both cases is excellent, with the almost imperceptible differences being caused by meshing differences and/or different considerations of the silicon permittivity (in the case of simulations with the disk). It should be mentioned that a major factor in the accuracy of this approach resides in the tiny refractive index contrast between water and glass, which is responsible for the constant behaviour of the low-angle Fresnel coefficients. This can be observed from the transmission and reflection factors, where reflection is energetically negligible ($\mathcal{R}_{TE,TM} < 0.005$). Due to this, the shape of the field around the silicon disk (Figure \ref{fig:S7})c-d) does not deviate heavily from the simulations in homogeneous media shown in Figure \ref{fig:1}e and \ref{fig:1}g in the main text.

\bibliography{ForceDisks.bib}

\end{document}